\documentclass[12pt,preprint]{aastex}

\newcommand{\dmm}{\mbox{$\Delta$m$_{15}(B)$}}

\newcommand{\ubvri}{\protect\hbox{$U\!BV\!RI$} }

\shorttitle{SN 2003hn}
\shortauthors{Krisciunas et al.}

\begin{document}

\title{Do the photometric colors of 
Type II-P Supernovae allow accurate 
determination of host galaxy extinction?\altaffilmark{1}}

\author{Kevin Krisciunas,\altaffilmark{2}
Mario Hamuy,\altaffilmark{3}
Nicholas B. Suntzeff,\altaffilmark{2}
Juan Espinoza,\altaffilmark{4}
David Gonzalez,\altaffilmark{4}
Luis Gonzalez,\altaffilmark{3}
Sergio Gonzalez,\altaffilmark{5}
Kathleen Koviak,\altaffilmark{6}
Wojtek Krzeminski,\altaffilmark{5}
Nidia Morrell,\altaffilmark{5}
Mark M. Phillips,\altaffilmark{5}
Miguel Roth,\altaffilmark{5} and
Joanna Thomas-Osip\altaffilmark{5}
}

\altaffiltext{1}{Based in part on observations taken at the Cerro Tololo 
Inter-American Observatory, National Optical Astronomy Observatory, which is 
operated by the Association of Universities for Research in Astronomy, Inc. (AURA) 
under cooperative agreement with the National Science Foundation. This paper 
also includes data gathered with the 6.5 meter Magellan Telescopes located at Las 
Campanas Observatory, Chile.
}

\altaffiltext{2}{Department of Physics, Texas A\&M University, 4242 TAMU,
College Station, TX 77843; {krisciunas@physics.tamu.edu}, {suntzeff@physics.tamu.edu} }

\altaffiltext{3}{Departmento de Astronom\'{i}a, Universidad de Chile,
Casilla 36D, Santiago, Chile; {mhamuy@das.uchile.cl}}

\altaffiltext{4}{Cerro Tololo Inter-American Observatory, Casilla 603,
  La Serena, Chile; {jespinoza@ctio.noao.edu} }

\altaffiltext{5}{Las Campanas Observatory, Casilla 601,
  La Serena, Chile; {sgonzalez@lco.cl}, {wojtek@lco.cl}, {nmorrell@lco.cl},
  {mmp@lcoeps1.lco.cl}, {mroth@lco.cl}, {jet@lco.cl} }

\altaffiltext{6}{Observatories of the Carnegie Institution of Washginton, 813 Santa Barbara 
  Street, Pasadena, CA 91101}

\begin{abstract}
We present infrared photometry of SN~1999em, plus optical photometry,
infrared photometry, and optical spectroscopy of SN~2003hn.  Both objects were 
Type II-P supernovae.  The $V-[RIJHK]$ color curves of these supernovae evolved 
in a very similar fashion until the end of plateau phase.
This allows us to determine how much more extinction the light of SN~2003hn suffered   
compared to SN~1999em.  Since we have an estimate of the
total extinction suffered by SN~1999em from model fits of ground-based and space-based
spectra as well as photometry of SN~1999em,
we can estimate the total extinction and absolute magnitudes of SN~2003hn
with reasonable accuracy.  Since the host galaxy of SN~2003hn also produced the Type Ia
SN~2001el, we can directly compare the absolute magnitudes of these two SNe of different
types.
\end{abstract}

\section{Introduction}

Supernovae (SNe) come in two basic models: exploding white
dwarfs that are members of close binary systems, and single massive stars that
develop iron cores. 

Type Ia SNe are generally regarded to be exploding carbon-oxygen white dwarf stars
that have reached the Chandrasekhar limit due to mass transfer from a nearby
non-degenerate donor star \citep[][and references therein]{Liv00}.  
Therefore, Type Ia SNe are explosions constrained by a
uniform energy budget.  This leads to a high degree of uniformity of their light
curves.  The objects with more rapidly declining light curves are fainter at maximum
light, and the slow decliners are brighter \citep{Phi93, Ham_etal96, Rie_etal96,
Per_etal97, Phi_etal99, Jha_etal07}. In the near-infrared, however, Type Ia SNe are
even better.  They are standard candles \citep{Kri_etal04a, Kri_etal04b, Woo_etal07}.  
Except for the fastest declining objects, there are essentially no ``decline rate
relations'' in the IR.

As we have shown in a recent series of papers \citep[see][and references
therein]{Kri_etal07,Wan_etal08}, a combination of optical and near-IR photometry allows the accurate
determination of host galaxy extinction of Type Ia SNe, even if the dust is different
than ``standard'' Galactic dust with R$_V$ = 3.1.

Type II SNe are thought to be single stars born with 8 M$_{\odot}$ or more which,
after a few million years of evolution, undergo the collapse of their iron cores and
the subsequent ejection of their hydrogen-rich envelopes \citep{Heg_etal03}.
Observationally, Type II-P SNe are distinguished by prominent hydrogen lines in their
spectra. When the star explodes with a significant fraction of its H-rich envelope, in
theory it should display a light curve characterized by a phase of $\sim$ 100 days of
nearly constant luminosity followed by a sudden drop of 2-3 mag \citep{Lit_etal83}.  
More than half of all Type II SNe belong to this class of ``Plateau'' SNe   
whose progenitors are attributed to stars born with less than 25 M$_{\odot}$
\citep{Hen_etal06,Li_etal07}. Type II-P SNe show a wide mass range, a considerable spread in
explosive power, absolute magnitudes, and ejecta velocities \citep{Ham_03}.  
\citet{Ham_Pin02} have shown that there is a correlation between the expansion
velocities of Type II-P SNe and their bolometric luminosities during the
plateau phase.  Thus, Type II-P SNe can be useful as standardizable candles in their
own right.

In this paper we address a simple question.  Can Type II-P SNe be found that exhibit
similar enough color curves such that we may attribute the systematic differences of
their colors to different amounts of dust extinction along the light of sight?  If
the answer to this question is Yes, then in principle photometry can be used to
obtain absolute magnitudes of Type II-P SNe which have minimal systematic errors
owing to dust extinction along the line of sight. Accurate distances give us
accurate luminosities, which are critical for constraining hydrodynamic models of
Type II-P SNe as well as for cosmological studies.

\section{SN 1999em}  

SN~1999em was discovered on 1999 October 29.44 by \citet{Li99} in the face-on spiral
NGC 1637. A spectrum taken on October 30.34 UT by \citet{Jha_etal99} revealed it to
be a Type II supernova at an early epoch.  \citet{Jha_etal99} give the correct
offsets of the SN from the core of its host galaxy (15.4 arcsec west and 17.0 arcsec
south). This object was intensively followed as part of the Supernova Optical and
Infrared Survey (SOIRS).\footnote[7]{The SOIRS project utilized several telescopes
in Chile and one in Arizona. See \citet{Ham_2001} for further details.}  The
Galactic reddening along the line of sight to SN~1999em is small, E($B-V$) = 0.040
\citep{Sch_etal98}, giving A$_V$(Gal) = 0.124 $\pm$ 0.012 mag.

A large fraction of the data of the two SNe discussed in this paper was obtained
with the dual optical-infrared photometer ANDICAM.  This instrument was mounted at
the Cassegrain focus of the Yale-AURA-Lisbon-Ohio (YALO) 1.0-m telescope at CTIO
from 1999 through early 2003.  After that it has been used with the 1.3-m ex-2MASS
telescope at CTIO.  


Optical and near-IR light curves of SN~1999em were first presented graphically by
\citet{Ham_etal01} and need not be reproduced here.  The optical photometry of
SN~1999em will be published by \citet{Ham_etal08} as part of an extensive study of
Type II-P SNe. \citet{Leo_etal02} present independent \ubvri photometry of this
object.  We note that the data of Leonard et al. were obtained with the Katzman
Automatic Imaging Telescope at Lick Observatory, {\em not} with any of the
telescopes at CTIO.  The photometric colors of SN~1999em as measured with the
various telescopes are in very good agreement.

Revised analysis of SN~1999em using the ``Expanding Photosphere Method'' (EPM) yields an
explosion time of JD 2,452,475 $\pm$ 1 day \citep{Jon_etal08}.  If all Type II-P SNe
could yield times of explosion with this accuracy, then these would be the most sensible
reference times.  \citet{Oli_etal08} uses the time that signals the end of the plateau.  
But it is known that the length of the plateau varies, as it depends on the mass of the
ejected envelope.  For our purposes here we shall adopt a ``reference time'' of Julian
Date 2,451,483.7 (= 1999 November 1.2 UT) for SN~1999em, which corresponds to a date
early in the photometric record. The plateau phase of the light curves lasted for roughly
100 days after this time.

The near-IR photometry of SN~1999em is presented in Table \ref{99em_jhk}.  The
calibration of the IR photometry was accomplished using observations of the
SN~1999em field and \citet{Per_etal98} standards on five photometric nights.

The $V-$[$IJHK$] colors of SN 1999em are shown in Fig. \ref{99em_colors}.
These will be our primary references for an investigation of the colors of
SN~2003hn.  Our color templates were obtained by fitting subsets of SN~1999em data with low
order polynomials, then stitching the fits together.  Until 100 days after the ``reference 
time'' the RMS scatter about these loci is less than $\pm$0.02 mag for $V-R$ and $V-I$, and 
is roughly $\pm$0.04 mag for $V-J$, $V-H$, and $V-K$.  

\section{SN 2003hn}  

SN~2003hn was discovered on 2003 August 25.7 UT by \citet{Eva03}.  The supernova was found
some 47 arcsec east and 53 arcsec north of
the nucleus of the side-on spiral galaxy NGC 1448 \citep{Kri_Esp03}.  This is the same galaxy
that hosted the Type Ia SN~2001el \citep{Kri_etal03, Kri_etal07}.  Using the 
{\sc iraf}\footnote[8]{{\sc iraf} is distributed by the National
Optical Astronomy Observatory, which is operated by the Association of Universities for
Research in Astronomy, Inc., under cooperative agreement with the National Science
Foundation.} tasks {\sc ccmap} and {\sc cctran} we find an accurate position of 
SN~2003hn to be RA = 3:44:36.27, DEC = $-$44:37:50.1 (J2000).

A finder chart is shown
in Fig. \ref{03hn_finder}.  A spectrum obtained with the 2.3-m telescope of the Australian
National University on 26.7 August UT by \citet{Sal_etal03} indicated that SN~2003hn was a
Type II-P supernova approximately two weeks after explosion (see Fig. \ref{03hn_stack}).

For the calibration of the optical photometry of SN~2003hn we adopt the magnitudes and
colors of the NGC 1448 field stars given by us previously \citep{Kri_etal03}.  From
further observations of this field on photometric nights in 2003 using \citet{Lan92}
standards we uncovered no anomalies in the optical calibration.  However, our previous IR
calibration of one of the field stars is wrong.  The star in question is labelled ``C2''
in Fig. \ref{03hn_finder}.  \citet{Kri_etal03} refer to it as ``star 7'' of their
photometric sequence.  From new observations on four photometric nights using infrared
standards of \citet{Per_etal98} we obtain the following photometry for ``star 7'': $J$ =
12.964 $\pm$ 0.007, $H$ = 12.500 $\pm$ 0.009, $K$ = 12.395 $\pm$ 0.025.  These can be
compared to the values from the Two Micron All Sky Survey (2MASS) of $J$ = 12.916 $\pm$
0.021, $H$ = 12.478 $\pm$ 0.021, $K$ = 12.439 $\pm$ 0.027.  Our $JHK$ photometry of
``star 6'' (= ``C3'' in Fig. \ref{03hn_finder}) obtained in 2003 is within a few
thousandths of a magnitude of the values published previously.

ANDICAM also contains a 1.03 $\mu$m filter known as $Y$ \citep{Hil_etal03}.
From synthetic photometry of Kurucz model spectra spanning a range of temperatures 
\citet[][Appendix C]{Ham_etal06} obtained a relationship between $Y-K_s$ colors and the
published $J-K_s$ colors of \citet{Per_etal98} standards. For ``star 6'' of the NGC 1448 
sequence we adopt Y = 14.102 $\pm$ 0.010, and for ``star 7'' Y = 13.279 $\pm$ 0.018.

Our optical photometry of SN~2003hn was derived using sharp reference templates obtained
with the CTIO 0.9-m telescope on 2002 February 19 (UT), long {\em before} the supernova
exploded. The Las Campanas optical data published separately by \citet{Ham_etal08} used
templates obtained on 2004 October 4 and 2004 November 11.  It is possible that some of
their late time photometry is affected by the (faint) presence of the SN in their
templates.

For our optical photometry we used image subtraction scripts provided to us by Brian
Schmidt which rely on the PSF matching technique of \citet{Ala_Lup98}. This provided us
with PSF magnitudes of the SN and the field stars.  We then completed the data reduction
in the {\sc iraf} environment using the {\sc photcal} package.

%
%
%

In Table \ref{03hn_ubvri} we present \ubvri photometry of SN~2003hn obtained at CTIO.
Near-IR photometry obtained with the CTIO 1.3-m telescope is given in Table
\ref{03hn_yjhk}.

Filter by filter photometry of SN~2003hn is shown in Fig. \ref{03hn_all}. Photometric
colors based solely on optical data of SN 2003hn are shown in Fig. \ref{03hn_opt}.  $V$
minus IR colors of SN 2003hn are shown in Fig. \ref{03hn_ir}.  Whereas the plateau phase
of the light curves of SN~1999em lasted roughly 100 days, the
plateau phase of SN~2003hn lasted about 20 days less.  The onset of the sudden rise in
many color indices at late times occurred about 20 days earlier in SN~2003hn.

\citet{Jon_etal08} have carried out an EPM analysis of the spectra of
SN~2003hn obtained at Las Campanas Observatory.  
Their derived explosion time is JD 2,452,857 $\pm$ 4 days. The large
uncertainty in the date of explosion is due to the lack of early-time spectra compared to
SN~1999em.  The method of \citet{Oli_etal08} relies on the end of the plateau phase as
the reference time. This is not the best reference time to choose for our purposes here,
as the plateau phase of SN~2003hn lasted 20 days less than that of SN~1999em. We
choose to adopt a ``reference time'' for SN~2003hn which relies on a $\chi^2$
minimization of the color curves using the templates from SN~1999em for fitting.  The
most consistent results were obtained using the $V-J$, $V-H$, and $V-K$ templates and
adding 1386.78 $\pm$ 0.45 days to the reference time adopted for SN~1999em.  Thus, for
SN~2003hn we obtain a reference time of JD 2,452,870.48 $\pm$ 0.45.

If two SNe have identical spectra but suffer different amounts of total extinction along
the line of sight, {\em all} color indices of one object should be redder than the
corresponding color indices of the other object.  In most color indices ($V-$[$RIJHK$])
SN~2003hn was redder than SN~1999em.  However, until the end of its plateau phase
SN~2003hn was {\em bluer} than SN~1999em in $U-B$ by 0.154 mag and bluer in $B-V$ by
0.019 mag. The simplest explanation is that the spectrum of SN~1999em exhibits greater
line blanketing due to a higher metallicity. Thus, we believe that these two objects do
not have identical spectra over the whole optical wavelength range. 

In Table \ref{color_excess} we give the magnitude shifts necessary to fit the SN~1999em
$V-$[$RIJHK$] color templates to the SN~2003hn colors.  Under the assumption that the
differences are solely due to dust extinction, these color excesses should monotonically
increase as the wavelength of the second filter increases. The reddening model of
\citet{Car_etal89} easily allows us to calculate multiplicative factors to scale these
color excesses, giving implied values of $\Delta$A$_V$ that should be statistically
equal. Column 6 of Table \ref{color_excess} gives these estimates and a weighted mean of
$\Delta$A$_V$ = 0.245 $\pm$ 0.025.  This was obtained assuming a value of R$_V$ = 3.1.
The small reduced $\chi^2$ values indicate that
the photometry may actually be more accurate than the formal random errors.
If we had used R$_V$ = 2.15, the value derived by \citet{Kri_etal07} for the host galaxy
dust affecting SN~2001el, we obtain a very similar value of $\Delta$A$_V$ = 0.225 $\pm$ 0.023
for the difference of the $V$-band extinctions of SNe 2003hn and 1999em.

In the third column of Table 4 we give the RMS scatter of the SN~2003hn
data with respect to the adjusted SN~1999em color templates.  These values
are comparable to the RMS scatter of the SN~1999em data used to create the templates.
Thus, the shape and precision of the two sets of color curves are comparable.

\citet{Bar_etal00} used the spectral synthesis code SYNOW to fit ground-based spectra and
one HST spectrum of SN~1999em obtained within a week of its discovery.  To match the
observed spectra with SYNOW model spectra required reddening the latter by E($B-V$) =
0.10, with an uncertainty of $\pm$ 0.05 mag.  Assuming ``standard'' R$_V$ = 3.1 dust
implies that the total $V$-band extinction suffered by SN~1999em was A$_V$ = 0.31 $\pm$
0.15 mag.  \citet{Oli_etal08} find A$_V$(host) = 0.24 $\pm$ 0.14, implying 
A$_V$(total) = 0.364 for SN~1999em.  We shall adopt A$_V$(total) = 0.34 $\pm$ 0.14 for 
SN~1999em.  

The implication is that the line of sight to SN~2003hn was affected by a total extinction
of A$_V$ = 0.58 $\pm$ 0.14 mag.  Only 0.043 mag of this extinction is due to dust in our
Galaxy, as E($B-V$)$_{Gal}$ = 0.014 towards NGC 1448 \citep{Sch_etal98}.
\citet{Oli_etal08} obtain A$_V$(total) = 0.50 $\pm$ 0.14 mag from a consideration of the
$V-I$ color only. At our request, \citet{Des08} kindly performed fits of Type II-P SN
atmosphere models to our optical spectra of SN~2003hn. In these fits the spectral lines
are used to constrain the photospheric temperature and the corresponding continuum is
employed to estimate the extinction. His analysis yields A$_V$(total) = 0.63 $\pm$ 0.25
mag.  These three estimates are remarkably consistent. It is more likely that our
optical/IR derived value is closest to the truth, as it relies on a large number of
photometric bands extending over a wider range of wavelengths than either of the other
two methods.  Also, since interstellar extinction is less problematic at IR wavelengths,
$V$ minus IR color excesses asymptotically approach A$_V$ as the wavelength of the IR band
under consideration becomes longer and longer.

In Table \ref{03hn_spectra} we summarize the available spectra of SN~2003hn. The first
spectrum was obtained with the 2.3-m telescope of the Australian National University and
the remaining nine spectra at Las Campanas Observatory. This spectroscopic time series,
shown in Fig. \ref{03hn_stack}, starts 21 days after explosion [assuming the time of
explosion on JD 2452857 $\pm$ 4 derived by \citet{Jon_etal08} using the ``Expanding
Photosphere Method''] and covers 163 days of evolution of SN~2003hn.  The strongest SN
lines are indicated along with the telluric lines (by the ``T'' symbol). The first
spectrum shows a blue continuum, P-Cygni profiles for the H Balmer lines, and a weak He I
$\lambda$5876 line which is characteristic of Type II SNe during their initial hottest
phases. The expansion velocity from the minimum of the H$\beta$ absorption feature was
$\sim$9,400 km~s$^{-1}$ which is typical of Type II SNe during the initial phases. The
presence of the interstellar Na I D lines $\lambda\lambda$5890, 5896 with an equivalent
width of 0.8$\pm$0.1~\AA~ suggest non-negligible interstellar absorption in the host
galaxy (in agreement with our analysis of the SN colors shown above). As the SN evolved
the atmospheric temperature dropped, the He I $\lambda$5876 line disappeared, and several
new lines became evident, namely, the Ca II H\&K $\lambda\lambda$3934,3968 blend, the Ca
II triplet $\lambda\lambda$8498,8542,8662, the Na I D blend, and several lines attributed
to Fe~II [see \citet{Jef_etal90} for a list of line identifications].  By day 90, the
color temperature was only 5,000 K, approximately the recombination temperature of H.
This spectrum corresponds to the end of plateau or the optically thick phase. The last
three spectra were taken during the nebular phase as can be appreciated from the
relatively fainter continuum and the increasingly weaker absorption features.

\section{Discussion}  

\citet{Ham_etal01} give an EPM distance to SN~1999em of 7.5 $\pm$ 0.5 Mpc. A similar
analysis by \citet{Leo_etal02} yields 8.2 $\pm$ 0.6 Mpc. Using improved photospheric
velocities \cite{Jon_etal08} obtain a distance of 9.3 $\pm$ 0.5 Mpc. All these EPM
distances were obtained using the same set of dilution factors calculated by
\citet{Eas_etal96}. Recently a new set of atmosphere models was calculated by
\citet{Des05} whose dilution factors are systematically higher than those of Eastman
et al. and which lead to greater distances. Based on these new models \citet{Des06} and
\citet{Jon_etal08} obtain 11.5 $\pm$ 1.0 and 13.9 $\pm$ 1.4 Mpc, respectively.
\citet{Leo_etal03} give a Cepheid-based distance of 11.7 $\pm$ 1.0 Mpc, in good agreement
with the EPM results derived from the new models of \citet{Des05}, which are
considerably greater than those obtained with the \citet{Eas_etal96} models. If we adopt
the Cepheid-based distance, the maximum observed $V$-band magnitude of 13.79 and total
$V$-band extinction of 0.34 mag, it follows that M$_V$ = $-$16.89 $\pm$ 0.24 mag for
SN~1999em.

Using the \dmm\ method of \citet{Phi_etal99}, the distance to NGC 1448, the host of SNe
2001el and 2003hn, is 17.9 $\pm$ 0.8 Mpc \citep{Kri_etal03}. Adopting the total $V$-band
extinction of 0.586 $\pm$ 0.050 mag for SN~2001el \citep{Kri_etal07}, it follows that
M$_V$(max) = $-$19.12 $\pm$ 0.11.  A check on the distance of NGC 1448 can be obtained by
assuming that the $JHK$ absolute magnitudes at maximum of SN~2001el equal the mean values
given in Table 17 of \citet{Kri_etal04b}.  \citet{Kri_etal07} also found that R$_V$ =
2.15 is the most appropriate value for the host galaxy dust associated with SN~2001el.  
From the near-IR maxima we obtain a distance of 18.1 $\pm$ 0.4 Mpc.  For comparison, the
EPM analysis of SN~2003hn by \citet{Jon_etal08} yields 16.9 $\pm$ 2 and 26.3 $\pm$ 7 Mpc,
using dilution factors from \citet{Eas_etal96} and \citet{Des05}, respectively. The
``Standardized Candle Method'' (SCM) applied to SN~2003hn yields a distance of 17.8 $\pm$
1 Mpc \citep{Oli_etal08}.

Under the assumption that the early-time photometric behavior of SN~2003hn was the same
as that of SN~1999em, we can extrapolate that SN~2003hn was 0.056 mag brighter than our
earliest $V$-band measurement.  Adopting V$_{max}$ = 14.41 $\pm$ 0.03, A$_V$(total) =
0.58 $\pm$ 0.14, and $d$ = 17.9 $\pm$ 0.8 Mpc, it follows that M$_V$(max) = $-$17.44
$\pm$ 0.17. Taken at face value, SN~2003hn was 0.55 $\pm$ 0.30 mag brighter than its
``cousin'' SN~1999em.  Since SN~2003hn and 2001el occurred in the same galaxy, a
comparison of their absolute magnitude differences involves no uncertainties in distance.  
Corrected for extinction, at the time of maximum light SN~2003hn was 1.68 mag fainter 
in $V$ than SN~2001el.  This confirms the notion that a typical Type II-P SN is 
significantly fainter than a Type Ia SN for the optical maxima.

In the near-IR, SNe 2001el and 2003hn are not so dissimilar in brightness at maximum.  
For SN~2001el the observed $K_{max}$ = 12.83 $\pm$ 0.04 \citep{Kri_etal03}.  Its total
$K$-band extinction was about 0.057 mag.  The absolute magnitude M$_K$(max) $\approx
-18.49$.  For SN~2003hn, $K_{max}$ = 13.27 $\pm$ 0.03 (from Table \ref{03hn_yjhk}), A$_K
\approx 0.064$ mag, so M$_K$(max) $\approx -18.06$.  This is only 0.43 mag fainter than
SN~2001el.  The implication is that a wide angle SN survey carried out at 2.2 microns
would find Type II-P SNe almost as easily as Type Ia SNe.

Since Type II SNe are single massive stars that have very short main sequence lifetimes,
they end their lives very close to where they were born, in regions of significant levels
of star formation.  As a result, the light curves of most (or all) Type II SNe would be
affected by interstellar extinction along the line of sight.

We have found two Type II-P SNe whose color curves vary in a similar manner
until the end of the plateau phase. The similarities of these color curves and
increasing color excesses as we proceed from $V-R$ through $V-K$ imply that the observed
color differences are simply due to differing amounts of dust along the line of sight.  
The implication is that some fraction of Type II-P SNe may exhibit sufficiently uniform color
curves that they may be used for a determination of the {\em relative} amounts of dust
extinction that they suffer.  Once we are confident we have data on Type II-P SNe which
are minimally reddened in their host galaxies, we can obtain accurate total extinctions
for all the SNe with good optical and IR light curves which show those similar color
curves. This will lead to a more accurate
distance calibration for Type II-P SNe.  As future projects such as Pan-STARRS and LSST
discover large numbers of SNe, we should be able to use Type Ia and Type II-P SNe for
cosmology.

\acknowledgments

The CTIO 1.3-m telescope is operated by the Small and Moderate Aperture Research
Telescope System (SMARTS) Consortium. We are particularly grateful for the scheduling
flexibility of SMARTS. ANDICAM was designed by Darren Depoy of Ohio State University.  
Without the flexibility of the operation of the 1.3-m telescope and the dual channel
design of ANDICAM, this paper could not have been written. We used data from the Two
Micron All-Sky Survey.  We thank Brian Schmidt for providing the classification spectrum
of SN~2003hn. M.H. acknowledges support provided by NASA through Hubble Fellowship grant
HST-HF-01139.01-A, by Fondecyt through grant 1060808, Centro de Astrof\'\i sica FONDAP
15010003, the Millennium Center for Supernova Science through grant P06-045-F funded by
``Programa Bicentenario de Ciencia y Tecnolog\'ia de CONICYT'' and ``Programa Iniciativa
Cient\'ifica Milenio de MIDEPLAN''.

\clearpage

\begin{deluxetable}{lcccc}
\tablewidth{0pt}
\tablecolumns{5}
\tablecaption{Near-Infrared Photometry of SN~1999em\label{99em_jhk}}
\tablehead{
\colhead{JD\tablenotemark{a}} &
\colhead{$J/J_s$} &
\colhead{$H$} &
\colhead{$K/K_s$} &
\colhead{Telescope\tablenotemark{b}}
}
\startdata
481.80 & 13.748	(0.015)	& 13.552 (0.015) & 13.349 (0.015) & 1 \\ 
482.69 & 13.544	(0.032)	& 13.413 (0.019) & 13.191 (0.038) & 2 \\ 
483.76 & 13.507	(0.015)	& 13.400 (0.015) & 13.205 (0.015) & 2 \\ 
483.78 & 13.620	(0.018)	& 13.288 (0.025) & 13.086 (0.029) & 1 \\ 
484.76 & 13.505	(0.015)	& 13.388 (0.015) & 13.200 (0.020) & 2 \\ 
485.73 & 13.486	(0.050)	& 13.373 (0.036) & 13.247 (0.069) & 2 \\ 
486.77 & 13.437	(0.015)	& 13.327 (0.015) & 13.149 (0.015) & 2 \\ 
487.75 & 13.445	(0.015)	& 13.338 (0.015) & 13.158 (0.015) & 2 \\ 
488.76 & 13.394	(0.015)	& 13.297 (0.015) & 13.132 (0.017) & 2 \\ 
489.81 & 13.378	(0.015)	& 13.269 (0.015) & 13.078 (0.015) & 2 \\ 
495.74 & 13.292	(0.015)	& 13.203 (0.015) & 13.142 (0.021) & 1 \\ 
498.68 & 13.342	(0.032)	& 13.247 (0.040) & 12.997 (0.033) & 1 \\ 
501.71 & 13.072	(0.015)	& 13.095 (0.015) & 12.875 (0.018) & 1 \\ 
504.74 & 13.289	(0.015)	& 13.059 (0.015) & 12.863 (0.017) & 1 \\ 
505.72 & 13.179	(0.015)	& 12.977 (0.015) & 12.823 (0.021) & 1 \\ 
507.80 & 13.158	(0.015)	& 13.041 (0.029) & 12.812 (0.031) & 1 \\ 
510.75 & 13.155	(0.015)	& 12.998 (0.015) & 12.845 (0.039) & 1 \\ 
513.72 & 13.089	(0.016)	& 12.959 (0.030) & 12.733 (0.027) & 1 \\ 
516.71 & 12.885	(0.015)	& 12.811 (0.015) & 12.665 (0.015) & 1 \\ 
519.72 & 12.981	(0.015)	& 12.844 (0.017) & 12.535 (0.015) & 1 \\ 
522.59 & 12.997	(0.015)	& 12.779 (0.015) & 12.658 (0.021) & 1 \\ 
527.63 & 12.942	(0.015)	& 12.770 (0.015) & 12.523 (0.016) & 1 \\ 
528.59 & 12.947	(0.015)	& 12.754 (0.015) & 12.544 (0.017) & 1 \\ 
538.60 & 12.894	(0.015)	& 12.710 (0.015) & 12.412 (0.015) & 1 \\ 
546.61 & 12.875	(0.015)	& 12.669 (0.015) & 12.514 (0.025) & 1 \\ 
547.60 & 12.843	(0.015)	& 12.687 (0.015) & 12.503 (0.017) & 1 \\ 
551.66 & 12.917	(0.015)	& 12.716 (0.015) & 12.552 (0.015) & 1 \\ 
558.55 & 12.828	(0.015)	& 12.730 (0.015) & 12.469 (0.015) & 1 \\ 
565.57 & 12.971	(0.015)	& 12.721 (0.015) & 12.538 (0.015) & 1 \\ 
572.54 & 12.944	(0.015)	& 12.842 (0.015) & 12.656 (0.038) & 1 \\ 
578.55 & 13.153	(0.015)	& 13.028 (0.015) & 12.854 (0.015) & 1 \\ 
586.52 &  \ldots	& 13.163 (0.015) & \ldots 	  & 1 \\ 
592.53 & 13.573	(0.015)	& 13.376 (0.035) & 13.095 (0.039) & 1 \\ 
599.52 & 13.992	(0.047)	& 13.633 (0.039) & 13.531 (0.038) & 1 \\ 
606.54 & 14.568	(0.040)	& 14.198 (0.043) & 14.041 (0.047) & 1 \\ 
613.53 & 14.619	(0.057)	& 14.341 (0.053) & 14.186 (0.064) & 1 \\ 
620.52 & 14.815	(0.042)	& 14.433 (0.047) & 14.151 (0.044) & 1 \\ 
627.50 & 14.778	(0.034)	& 14.513 (0.045) & 14.271 (0.029) & 1 \\ 
634.49 & 14.922	(0.038)	& 14.618 (0.051) & 14.400 (0.038) & 1 \\ 
638.51 & 14.886	(0.045)	& \ldots         &   \ldots       & 2 \\ 
639.49 & 14.860	(0.055)	& 14.585 (0.036) &   \ldots	  & 2 \\ 
640.49 & 14.898	(0.057)	& 14.603 (0.041) &   \ldots	  & 2 \\ 
641.48 & 14.888	(0.059)	&    \ldots	 & 14.521 (0.087) & 2 \\ 
641.50 & 14.881	(0.023)	& 14.677 (0.035) & 14.384 (0.033) & 1 \\ 
642.48 &    \ldots	& 14.651 (0.034) & 14.568 (0.084) & 2 \\ 
643.49 & 14.934	(0.041)	&   \ldots	 & 14.607 (0.050) & 2 \\ 
644.48 & 14.938	(0.056)	& 14.656 (0.037) &	   \ldots & 2 \\ 
650.48 & 15.017	(0.048)	&   \ldots	 &  \ldots	  & 2 \\ 
653.48 & 15.064	(0.019)	& 14.908 (0.028) & 14.654 (0.038) & 1 \\ 
655.48 & 15.098	(0.047)	& 14.958 (0.023) & 14.779 (0.029) & 1 \\ 
656.48 & 15.200	(0.020)	& 15.015 (0.021) & 14.818 (0.033) & 1 \\ 
670.46 & 15.384	(0.020)	&   \ldots       &   \ldots	  & 1 \\ 
\enddata
\tablenotetext{a} {Julian Date minus 2,450,000.}
\tablenotetext{b} {1 = CTIO 1.0-m (YALO) using $JHK$ filters;
2 = LCO 1.0-m using $J_s H K_s$ filters.}
\end{deluxetable}

\begin{deluxetable}{lcccccc}
\tablewidth{0pt}
\tabletypesize{\scriptsize}
\tablecolumns{7}
\tablecaption{Optical Photometry of SN~2003hn\label{03hn_ubvri}}
\tablehead{
\colhead{JD\tablenotemark{a}} &
\colhead{$U$} &
\colhead{$B$} &
\colhead{$V$} &
\colhead{$R$} &
\colhead{$I$} &
\colhead{Telescope\tablenotemark{b}}
}
\startdata
2877.82 & 13.941 (0.011) &  14.642 (0.031) & 14.464 (0.038) & 14.163 (0.031) & 14.078 (0.036) & 1 \\
2878.79 & 13.953 (0.041) &  14.603 (0.026) & 14.435 (0.026) & 14.175 (0.012) & 14.046 (0.026) & 2 \\
2879.80 & 14.079 (0.010) &  14.703 (0.037) & 14.499 (0.043) & 14.163 (0.037) & 14.086 (0.037) & 1 \\
2883.82 & 14.543 (0.018) &  14.857 (0.037) & 14.468 (0.033) & 14.143 (0.037) & 14.060 (0.034) & 1 \\
2891.81 & 15.619 (0.032) &  15.323 (0.031) & 14.637 (0.038) & 14.231 (0.037) & 14.077 (0.024) & 1 \\
2899.82 & 16.611 (0.038) &  15.692 (0.032) & 14.805 (0.036) & 14.373 (0.038) & 14.186 (0.042) & 1 \\
2905.85 & 16.797 (0.045) &  15.906 (0.026) & 14.856 (0.029) & 14.407 (0.032) & 14.222 (0.033) & 1 \\
2905.88 &   \ldots       &  15.881 (0.018) & 14.903 (0.014) & 14.448 (0.015) & 14.185 (0.023) & 2 \\
2911.80 & 17.099 (0.063) &  16.137 (0.029) & 14.978 (0.041) & 14.465 (0.033) & 14.223 (0.028) & 1 \\
2918.78 & 17.262 (0.071) &  16.256 (0.025) & 15.043 (0.034) & 14.547 (0.033) & 14.279 (0.032) & 1 \\
2925.75 & 17.620 (0.103) &  16.415 (0.034) & 15.116 (0.038) & 14.606 (0.042) & 14.306 (0.030) & 1 \\
2933.73 & 17.984 (0.143) &  16.563 (0.031) & 15.192 (0.040) & 14.631 (0.029) & 14.374 (0.042) & 1 \\
2939.86 & 18.204 (0.037) &  16.663 (0.021) & 15.282 (0.019) & 14.702 (0.017) & 14.375 (0.016) & 2 \\
2940.75 & 18.476 (0.235) &  16.744 (0.034) & 15.249 (0.038) & 14.722 (0.039) & 14.402 (0.029) & 1 \\
2947.68 &   \ldots       &  16.917 (0.039) & 15.400 (0.035) & 14.786 (0.040) & 14.527 (0.043) & 1 \\
2954.69 &   \ldots       &  17.210 (0.027) & 15.581 (0.032) & 14.950 (0.039) & 14.614 (0.031) & 1 \\
2962.66 &   \ldots       &  17.990 (0.039) & 16.220 (0.034) & 15.440 (0.029) & 15.098 (0.049) & 1 \\
2967.75 & 20.681 (0.171) &  18.881 (0.025) & 17.134 (0.020) & 16.208 (0.015) & 15.756 (0.017) & 2 \\
2969.65 &   \ldots       &  19.219 (0.200) & 17.405 (0.100) & 16.364 (0.035) & 15.964 (0.036) & 1 \\
2990.63 &   \ldots       &  19.571 (0.080) & 17.749 (0.060) & 16.678 (0.028) & 16.299 (0.035) & 1 \\
2997.69 &   \ldots       &  19.611 (0.080) & 17.840 (0.060) & 16.811 (0.035) & 16.398 (0.035) & 1 \\
3030.66 &   \ldots       &  19.896 (0.045) & 18.202 (0.030) & 17.193 (0.022) &   \ldots       & 2 \\
\enddata
\tablenotetext{a} {Julian Date {\em minus} 2,450,000.}
\tablenotetext{b} {1 = CTIO 1.3-m; 2 = CTIO 0.9-m.}
\end{deluxetable}

\begin{deluxetable}{lcccc}
\tablewidth{0pt}
\tablecolumns{5}
\tablecaption{Near-Infrared Photometry of SN~2003hn\label{03hn_yjhk}}
\tablehead{
\colhead{JD\tablenotemark{a}} &
\colhead{$Y$} &
\colhead{$J$} &
\colhead{$H$} &
\colhead{$K$} 
}
\startdata
2877.82 & 13.976 (0.031) &    13.886 (0.010) &  13.670 (0.017) &     \ldots      \\
2879.80 & 13.987 (0.020) &    13.820 (0.010) &  13.663 (0.011) &  13.481 (0.030) \\
2883.82 & 13.914 (0.022) &    13.739 (0.013) &  13.570 (0.014) &  13.333 (0.033) \\
2891.81 & 13.884 (0.020) &    13.687 (0.010) &  13.531 (0.011) &  13.310 (0.027) \\
2899.79 & 13.949 (0.020) &    13.724 (0.009) &  13.557 (0.011) &  13.318 (0.028) \\
2905.85 & 13.955 (0.020) &    13.720 (0.009) &  13.559 (0.011) &  13.358 (0.027) \\
2911.79 & 14.021 (0.020) &    13.773 (0.009) &  13.571 (0.011) &  13.384 (0.029) \\
2918.77 & 14.062 (0.021) &    13.790 (0.010) &  13.589 (0.011) &  13.274 (0.030) \\
2925.75 & 14.088 (0.020) &    13.830 (0.009) &  13.634 (0.011) &  13.279 (0.032) \\
2933.73 & 14.186 (0.020) &    13.871 (0.010) &  13.684 (0.011) &  13.369 (0.032) \\
2940.75 & 14.198 (0.020) &    13.914 (0.010) &  13.706 (0.011) &  13.414 (0.031) \\
2947.68 & 14.302 (0.021) &    14.002 (0.010) &  13.774 (0.012) &  13.489 (0.030) \\
2954.69 & 14.394 (0.021) &    14.138 (0.011) &  13.930 (0.012) &  13.716 (0.036) \\
2969.65 & 15.707 (0.049) &    15.231 (0.043) &  14.787 (0.035) &     \ldots      \\
2990.62 & 16.152 (0.035) &    15.727 (0.020) &  15.385 (0.024) &  15.334 (0.094) \\
2997.69 & 16.262 (0.037) &    15.845 (0.026) &  15.540 (0.027) &  15.505 (0.106) \\
\enddata
\tablenotetext{a} {Julian Date minus 2,450,000.}
\end{deluxetable}

\begin{deluxetable}{ccccccc}
\tablewidth{0pt}
\tablecolumns{7}
\tablecaption{Color Excesses and Implied Values of 
$\Delta$A$_V$\tablenotemark{a}\label{color_excess}}
\tablehead{
\colhead{Color Excess} &
\colhead{Value (mag)} &
\colhead{RMS (mag)} &
\colhead{Range\tablenotemark{b}} &
\colhead{$\chi^{2}_{\nu}$} &
\colhead{Factor\tablenotemark{c}} &
\colhead{$\Delta$A$_V$} 
}
\startdata

$\Delta$($V-R$)  &    0.063 & $\pm$0.026  & 7$-$84 & 0.84 & 4.016 & 0.253 (0.104)  \\   
$\Delta$($V-I$)  &    0.098 & $\pm$0.032  & 7$-$77 & 0.58 & 1.919 & 0.188 (0.061)  \\   
$\Delta$($V-J$)  &    0.184 & $\pm$0.033  & 7$-$84 & 0.68 & 1.393 & 0.256 (0.046)  \\   
$\Delta$($V-H$)  &    0.190 & $\pm$0.035  & 7$-$84 & 0.81 & 1.235 & 0.235 (0.043)  \\   
$\Delta$($V-K$)  &    0.267 & $\pm$0.055  & 7$-$84 & 1.33 & 1.129 & 0.301 (0.062)  \\
Weighted Mean    &          &        &        &      &       & 0.245 (0.025)  \\
   
\enddata
\tablenotetext{a} {Here we determine the implied amount of extra $V$-band extinction suffered
by SN~2003hn compared to SN~1999em.}
\tablenotetext{b} {Range in days since the ``reference time'' of SN~2003hn, JD = 
2,452,870.48.}
\tablenotetext{c}{Scale factors derived from values of A$_{\lambda}$/A$_V$ in Table 
3 of \citet{Car_etal89}, used to obtain A$_V$ from color excess.  It is assumed that R$_V$ = 
3.1.}
\end{deluxetable}

\begin{deluxetable}{ccrcccc}
\tablewidth{0pt}
\tablecolumns{7}
\tablecaption{Log of SN~2003hn Spectra \label{03hn_spectra}}
\tablehead{
\colhead{JD\tablenotemark{a}} &
\colhead{UT Date} &
\colhead{$\Delta$T\tablenotemark{b}} &
\colhead{Telescope\tablenotemark{c}} &
\colhead{$\lambda$ Range (\AA)} &
\colhead{$\Delta \lambda$ (\AA)} &
\colhead{Exptime(sec)}
}
\startdata
2878.20 & 2003 Aug 26 & 21  & 1 & 3433-9162 & 1.1/2.1 & 600$\times$2 \\
2897.86 & 2003 Sep 15 & 41  & 2 & 3600-9000 & 4.2 & 90 \\
2900.87 & 2003 Sep 18 & 44  & 3 & 3780-7280 & 2.0 & 300 \\
2908.83 & 2003 Sep 26 & 52  & 3 & 3800-9325 & 3.1 & 600 \\
2928.83 & 2003 Oct 16 & 72  & 3 & 3800-9325 & 3.1 & 600 \\
2948.83 & 2003 Nov 05 & 92  & 2 & 3600-9000 & 4.2 & 150 \\
2966.81 & 2003 Nov 23 & 110 & 3 & 3800-9325 & 3.1 & 600 \\
2989.75 & 2003 Dec 16 & 133 & 3 & 3800-9325 & 3.1 & 600 \\
2996.68 & 2003 Dec 23 & 140 & 3 & 3800-9325 & 3.1 & 600 \\
3040.69 & 2004 Feb 05 & 184 & 2 & 3600-9000 & 4.2 & 600 \\
\enddata
\tablenotetext{a} {Julian Date minus 2,450,000.}
\tablenotetext{b}{Number of days since the derived explosion time
of JD 2,452,857 $\pm$ 4 \citep{Jon_etal08}.}
\tablenotetext{c} {1 = Australian National University 2.3-m; 2 = Magellan \#2
(6.5-m Clay Telescope); 3 = 2.5-m DuPont Telescope.} 
\end{deluxetable}

\clearpage

\figcaption[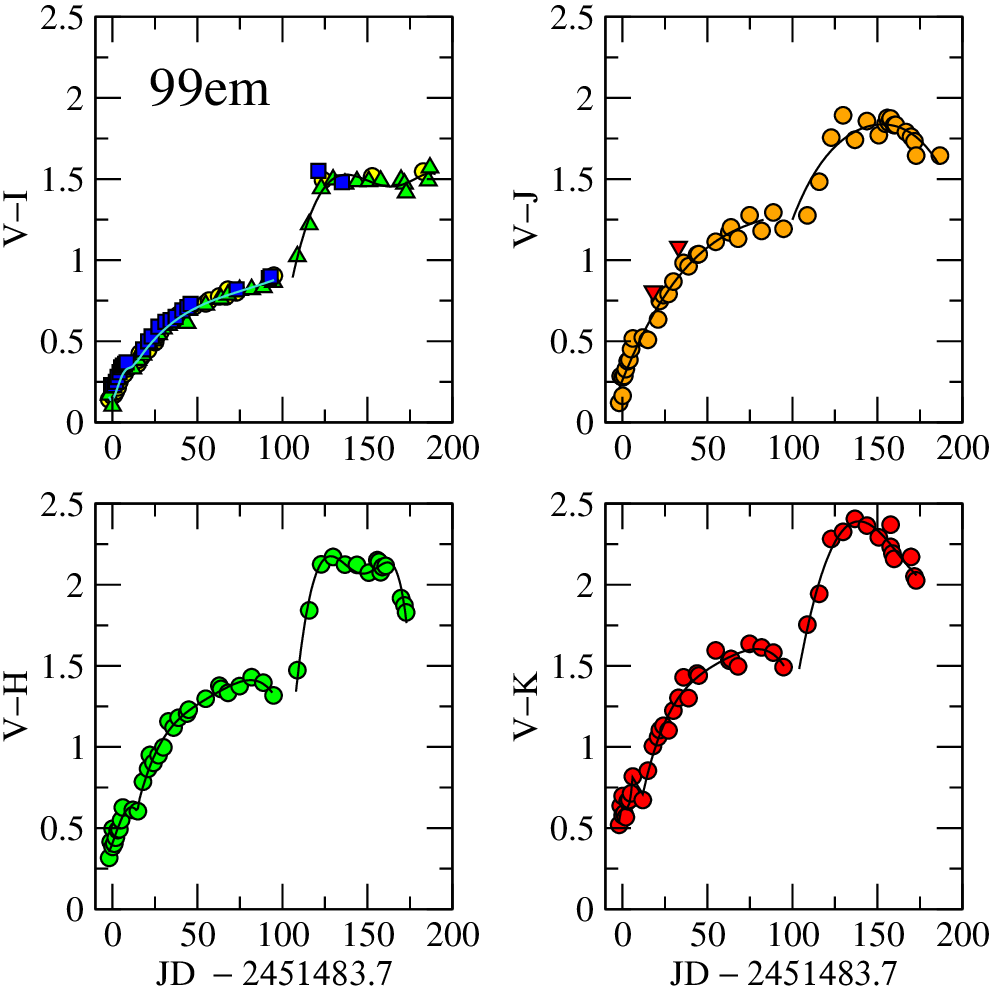] {Optical/IR colors of SN 1999em vs.
number of days since reference time, JD 2,451,483.7.  
In addition to the near-IR data presented in this
paper, we have used optical data of \citet{Leo_etal02} and \citet{Ham_etal08}.
The $V-R$ template of SN~1999em excludes $R$-band taken with ANDICAM.
At that time ANDICAM had a very broad, non-standard $R$-band filter.  
We have fit lower order polynomials to 
sections of the data.  Two points in the upper right diagram, marked by
triangles, have been treated as outliers and were excluded from the fits.
\label{99em_colors}
}

\figcaption[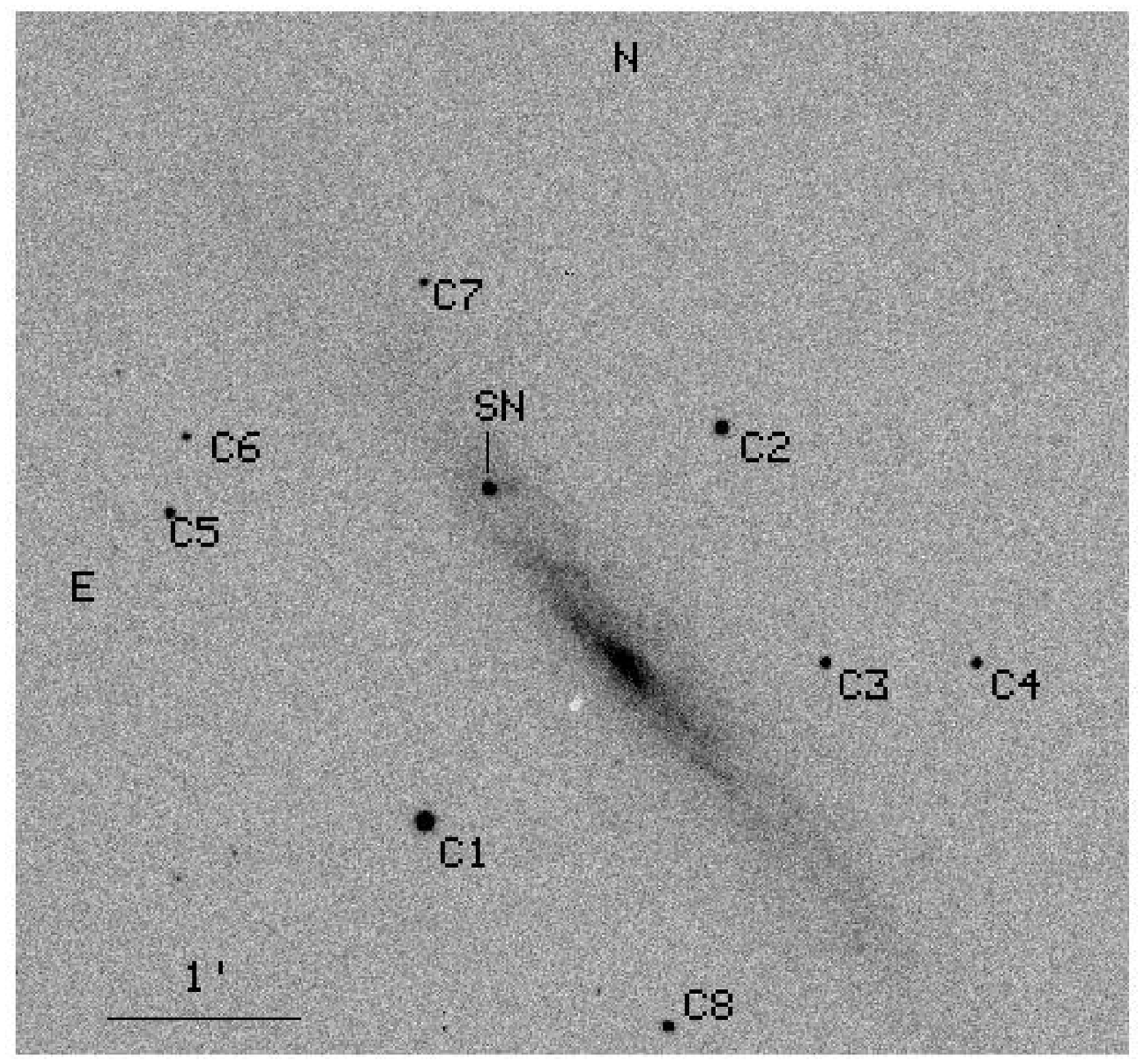] {Finder chart for NGC 1448 and
SN~2003hn.  This is a $V$-band image obtained with the Las Campanas 1.0-m
telescope on 2003 August 28 UT. The stars ``C2'' and ``C3'' here are the same 
as ``star 7'' and ``star 6,'' respectively, of the photometric sequence 
of \citet{Kri_etal03}. 
\label{03hn_finder}
}

\figcaption[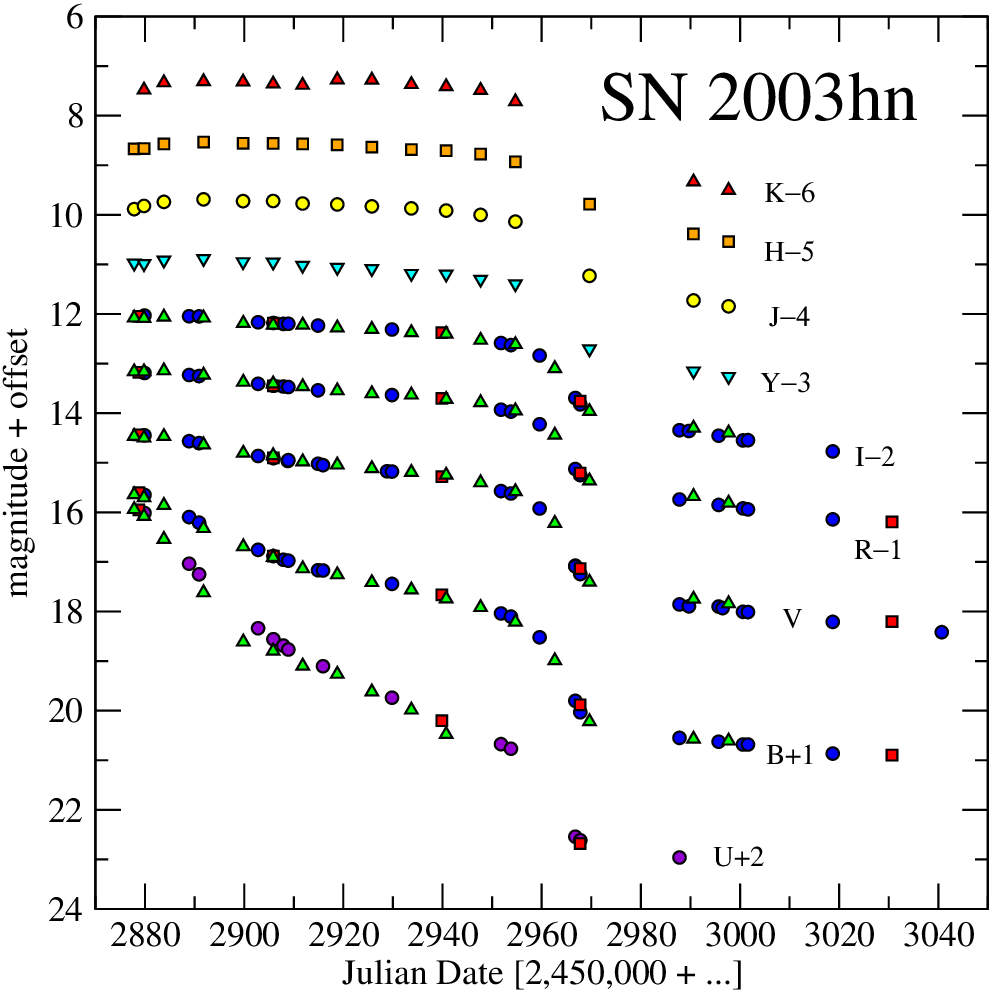] {Optical and near-IR photometry of SN~2003hn.
The \ubvri data are coded by telescope.  Dots = data from
Las Campanas Observatory 1.0-m; squares = CTIO 0.9-m; triangles = CTIO 1.3-m.
All the infrared data were obtained with the CTIO 1.3-m.  Except for some of
the late-time $U$- and $B$-band data, the uncertainties of the data points are 
comparable to, or smaller than, the size of the symbols.
\label{03hn_all}
}

\figcaption[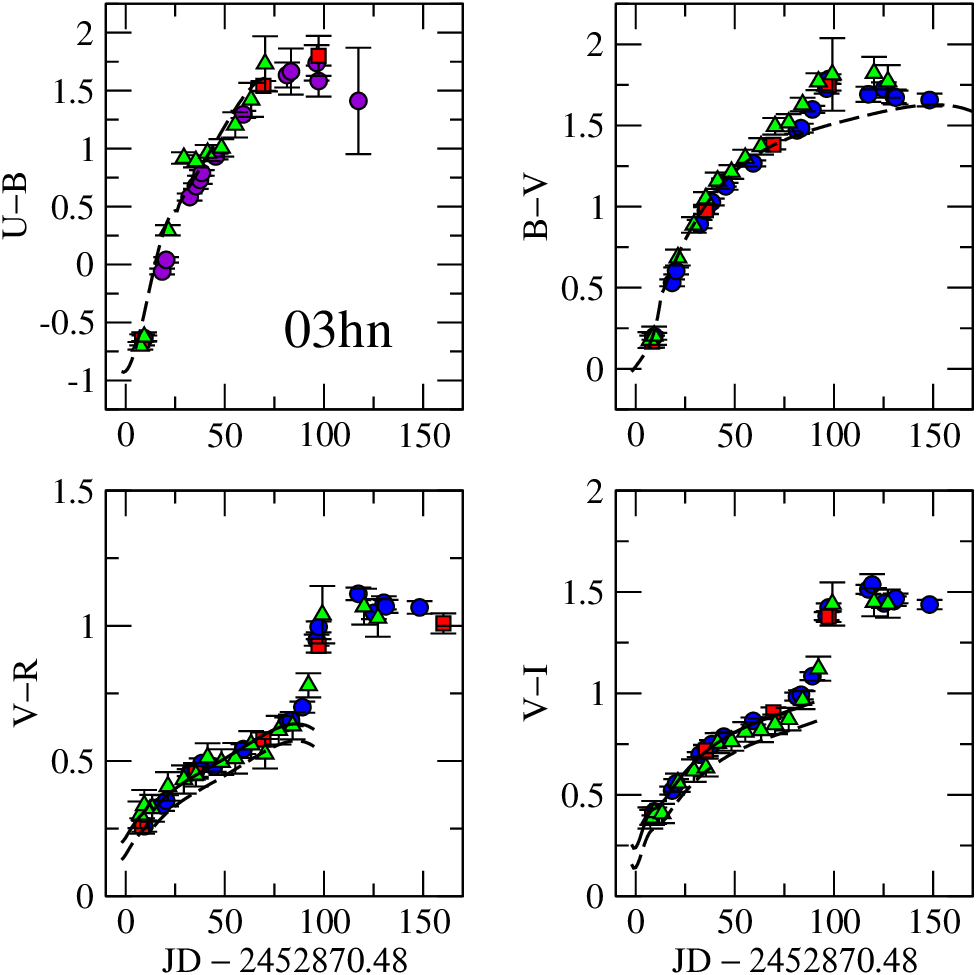] {Photometric colors of SN 2003hn based
solely on optical photometry.  The abscissa represents the number of days
since the reference time JD 2,452,870.48.  The dashed curves are the loci from 
SN~1999em, using data from \citet{Leo_etal02} and \citet{Ham_etal08}.  
The solid lines for $V-R$ and $V-I$ are the loci from 
SN 1999em offset by amounts that minimize the reduced $\chi^2$ of the
fits.
\label{03hn_opt}
}

\figcaption[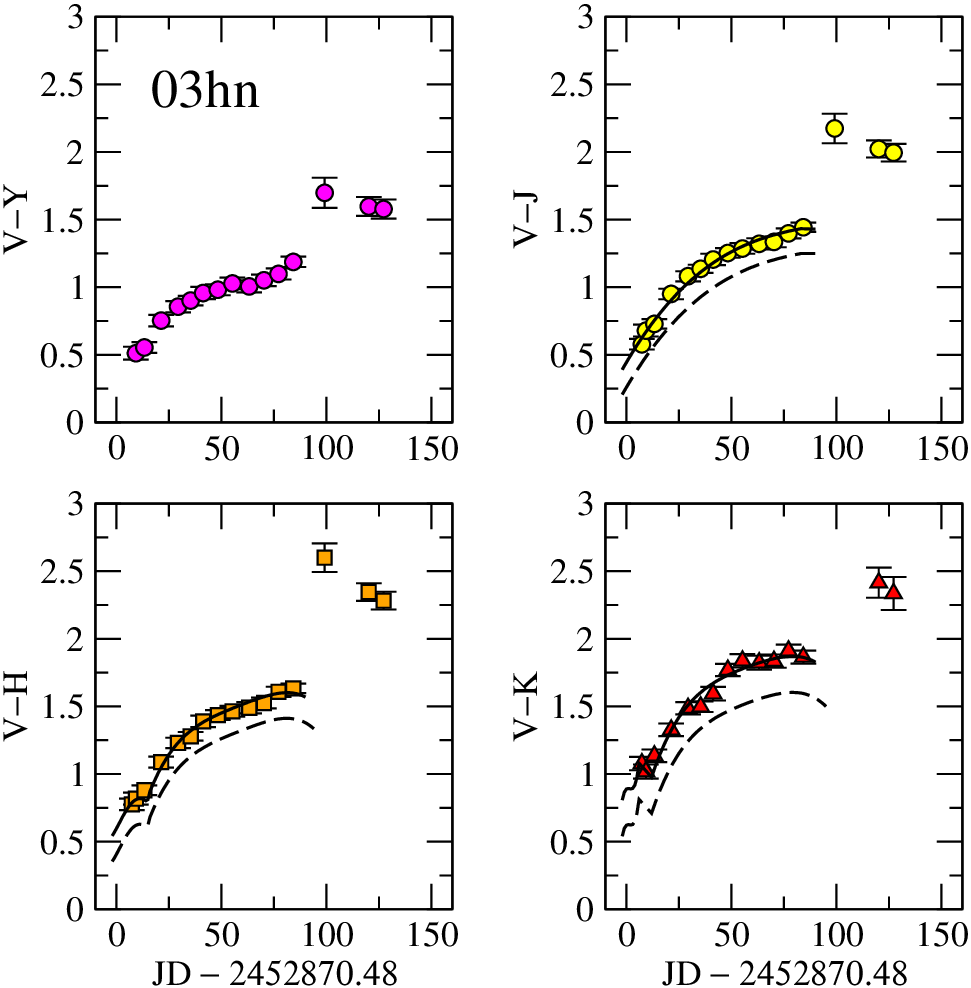] {$V$ minus near-IR colors of SN 2003hn
vs. days since the reference time JD 2,452,870.48.  As in Fig. \ref{03hn_opt}, the dashed
lines are the loci from SN~1999em, and the solid lines are those loci
offset to minimize the reduce $\chi^2$ of the fits to the data of
SN~2003hn.
\label{03hn_ir}
}

\figcaption[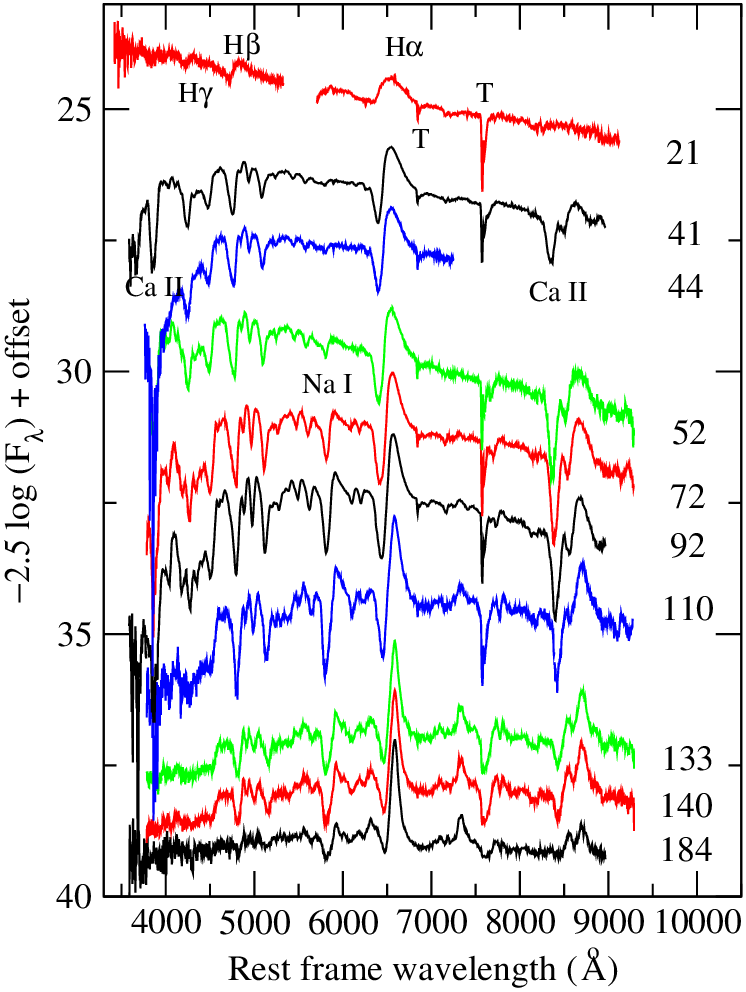] {Spectra of SN~2003hn.  See Table \ref{03hn_spectra}
for further details.  The numbers within the graph are the number of
days since the time of explosion, JD 2,452,857 $\pm$ 4 \citep{Jon_etal08}.
The features labelled ``T'' are terrestrial features due to the Earth's
atmosphere.
\label{03hn_stack}
}

\begin{figure}
\plotone{99em_v_ijhk.eps}
{\center Krisciunas {\it et al.} Fig. \ref{99em_colors}}
\end{figure}

\begin{figure}
\plotone{sn2003hn_chart.ps}
{\center Krisciunas {\it et al.} Fig. \ref{03hn_finder}}
\end{figure}

\begin{figure}
\plotone{03hn_all.eps}
{\center Krisciunas {\it et al.} Fig. \ref{03hn_all}}
\end{figure}

\begin{figure}
\plotone{03hn_opt_colors.eps}
{\center Krisciunas {\it et al.} Fig. \ref{03hn_opt}}
\end{figure}

\begin{figure}
\plotone{03hn_ir_colors.eps}
{\center Krisciunas {\it et al.} Fig. \ref{03hn_ir}}
\end{figure}

\begin{figure}
\plotone{stack.eps}
{\center Krisciunas {\it et al.} Fig. \ref{03hn_stack}}
\end{figure}

\end{document}